# Modeling the IPv6 Internet AS-level Topology[☆]

Bo Xiao, Lian-dong Liu, Xiao-chen Guo, and Ke Xu[*]

State Key Laboratory of Software Development Environment,
School of Computer Science and Engineering, Beihang University, Beijing 100083, China

**Abstract**

To measure the IPv6 internet AS-level topology, a network topology discovery system, called Dolphin, was developed. By comparing the measurement result of Dolphin with that of CAIDA's Scamper, it was found that the IPv6 Internet at AS level, similar to other complex networks, is also scale-free but the exponent of its degree distribution is 1.2, which is much smaller than that of the IPv4 Internet and most other scale-free networks. In order to explain this feature of IPv6 Internet we argue that the degree exponent is a measure of *uniformity* of the degree distribution. Then, for the purpose on modeling the networks, we propose a new model based on the two major factors affecting the exponent of the EBA model. It breaks the lower bound of degree exponent which is 2 for most models. To verify the validity of this model, both theoretical and experimental analyses have been carried out. Finally, we demonstrate how this model can be successfully used to reproduce the topology of the IPv6 Internet.

## 1. Introduction

As one of the most significant inventions in the last century, the Internet has provided human beings a brand new information society. Now the internet is undergoing a gradual change. As core parts of Internet, TCP/IP protocol stacks perform the tasks of packaging and transmitting information. On the IP layer, the Internet Protocol version 4 (IPv4) has the limitation of address space and some other problems such as the QoS, security and performance. Under this circumstance, the Internet Protocol version 6 (IPv6) has been put into practical use, leading to the coexistence of the IPv6 Internet and the IPv4 Internet. Currently, the IPv6 networks usually connect to each other through IPv4 tunnels. For the purpose of predicting how new technologies, policies, or economic conditions will impact the Internet's connectivity structure at different layers, the topology of global IPv6 Internet is necessary. It made use of those tunnels and treated them as the links between IPv6 domains[1-4]. The research can be either at router level or at autonomous system (AS) level. In fact, more interest boomed in AS topologies, because from a macroscopic view AS topologies are the "skeletons" of this complex system and more representative. The topology model is the outcome of theoretically modeling the real networks from the view of systematical evolution or the aspect of reproducing some important topology metrics [5, 6]. Therefore, topology modeling can explain the origins of the existing properties of network topologies and the model also contributes to research on network simulations and structural analysis.

In recent years, considerable research has been done on complex networks which describe a wide variety of systems in nature and society including Internet, World Wide Web (WWW), social relationship networks, economy networks, power networks, transportation networks and neural networks [7]. It is also known that some of the networks can be represented as scale-free networks[8], whose degree distribution follow the power-law form $p(k) \propto k^{-\gamma}$ where $p(k)$ is the probability that a randomly selected node has exactly $k$ edges and $\gamma$ is called the degree exponent which characterizes the degree distribution of a scale-free network. To understand the evolving mechanisms of scale-free networks, a number of evolving topology models have been proposed. A simple model of a growing network was introduced by Barabási and Albert (the BA model [8]) in which they found that the growth of networks and the preferential attachment were the origins of the power-law degree distribution and proposed the concept of "scale-free" networks. The BA model produces networks with the degree exponent $\gamma = 3$. Based on the BA model, a lot of other evolving models have been introduced to obtain degree distributions with variable degree exponents [9-18].

☆This research is supported by National 973 Program of China (Grant NO 2005CB321901) and Beijing Nova Program (Grant. NO 2005B12).
* Corresponding author. Tel.: +86 10 8231 5704; E-mail: kexu@nlsde.buaa.edu.cn

For many observed scale-free networks with $2 < \gamma < 3$, these models fit their degree exponent feature very well. However, a major challenge arises when using these models to reproduce the IPv6 Internet topology with quite a small degree exponent because most of them have the limitation of $\gamma > 2$ [7]. In this paper we propose our model to break this limitation and reproduce the IPv6 Internet AS-level topology.

The paper is organized as follows. In section 2, we first review the evolving models for network topologies. And then we briefly introduce the IPv6 Internet topology discovery and present a new feature of the AS-level topology in section 3. The degree exponent and its implications are discussed in section 4. Then, based on the two major factors affecting the exponent and the EBA model, we propose our model and make a theoretical analysis in section 5. It is shown in section 6 that our model breaks the bound of 2 for the degree exponent $\gamma$, and reproduces the topology of the IPv6 Internet. Finally, we conclude our work in section 7.

## 2. Evolving models for network topologies

In 1999, several power-laws were observed in the IPv4 Internet topology [19, 20]. The discovery of the power-law of degree distribution has significant impact on the network topology research: the Internet is neither as "flat" as the random-models (such as the ER model [21, 22]) describe, nor as "hierarchical" as the structural ones (such as the Tires and Transit-Stub models [23, 24]) describe. To model this scale-free property in the Internet, many efforts have been brought forward to design topology models that can produce power-law degrees. They all construct random graphs, but some produce the graphs incrementally, which are called *evolving models*; and others do not allow the growth of the network, which are called *non-evolving models* [25]. The non-evolving models can not satisfy our demands because we want to predict the future topology but these models only reproduce the same size as the given one. So we mainly discuss and compare the evolving models.

Barabási and Alert found that increase of networks and preferential attachment were the origins of the degree distribution power-law. They proposed the conception of scale-free networks and a linear preferential attachment model (the BA model) [8] to generate the scale-free networks. The BA model contributed a lot to complex network theory by finding the origins of the well-known power-law characteristic. In the BA model, the preferential attachment probability is defined as:

$$\Pi_i = \frac{k_i}{\sum_j k_j}. \tag{1}$$

However the BA model can only construct network topology with the degree exponent equal to 3. Then, Albert and Barabási improved their BA model and proposed the EBA model [9]. The new model is still a linear preferential model, but takes into account local events, such as suspension of adding nodes and rewiring of links. The distribution of degrees in the EBA model follows the power-law with the exponent $\gamma > 2$.

The Generalized Linear Preference (GLP) model [10] was proposed in 2002 by Bu and Towsley who discovered that the nodes in Internet are more likely to attach to the node with larger degree than the probability specified in the BA model. So they modified the preferential attachment probability to

$$\Pi_i = \frac{k_i - \beta}{\sum_j k_j - \beta}, \tag{2}$$

where $\beta < 1$. The degree distribution of GLP also follows the power-law with the exponent $\gamma > 2$.

Zhou and Mondragon argued that an Internet model that did not reproduce the properties of the *rich-club connectivity* would underestimate the actual network's routing efficiency and routing flexibility and it would also overestimate the network robustness under node attack [26]. The Positive Feedback Preference (PFP) model [11] has successfully reproduced a wide spectrum of metrics of the IPv4 Internet at AS-level, including the rich-club connectivity.

Other models in this category include the highly optimized tolerance (HOT) Model [12], the initial attractiveness model [33], the accelerating growth model [17] and the gradual aging model [17]. However, all of these models discussed above have the limitation of $\gamma > 2$, as shown later in section 4. In this paper, we will only choose BA and EBA, the two most basic models, for discussion and comparison.

## 3. The small degree exponent: a new feature of the IPv6 Internet topology

In this section we first analyzed the topology data from CAIDA's Scamper and Dolphin. We found that the IPv6 Internet at AS level, similarly with other complex networks, is also scale-free but its exponent of the degree distribution is 1.2, which is much smaller than that of the IPv4 Internet and most other scale-free networks.

3.1 Topology discovery on the IPv6 Internet

Now the global IPv6 networks are connected to each other mainly through IPv4 tunnels and can be seen as an independent network. Knowledge of this new born network topology is crucial for understanding current applications on IPv6 networks and identifying the problems when these applications are deployed. The IPv6 internet topology, as the "skeleton" of the communication system, is of great importance for testing and predicting the performance, robustness, and scalability of other network protocols on this new IP layer. And the topology can be helpful for the studies about routing and searching in networks, robustness to random network failures and targeted attacks, the speed of worms spreading, common strategies for traffic engineering and network management [27, 28].

In 1998, the Cooperative Association for Internet Data Analysis (CAIDA) [29] began the "Macroscopic Topology Project" for Internet topology measurements. CAIDA has developed a tool named *Skitter,* which is designed for the IPv4 Internet topology discovery and other measurements. Its counterpart on IPv6 is *Scamper* [30] which extends *Skitter*'s approach to the IPv6 protocol. *Skitter* and *Scamper* are both based on traceroute. For IPv6 networks, traceroute obtains the IP hops along the path from the probing point to a given destination by the ICMPv6 protocol. In order to measure the global internet, distribution is required. Unlike the distributed project DIMES[35], which realizes distribution by volunteers, Scamper was deployed at 17 different locations around the world and then began to collect data. A snapshot of the IPv6 Internet AS topology on March 4th, 2005 was published on the website [31].This AS map came from the discovery of approximately 860 globally routable IPv6 networks prefixed by 2,913 IPv6 addresses and 7,905 IPv6 IP links which are the links between adjacent addresses in a traceroute path, as shown in TABLE 1. In early 2007, CAIDA gave network researchers free access to the raw data of Skitter and Scamper after registration. There is a little difference between the AS graphs generated from the data we downloaded from CAIDA and the one from the site [31]: the AS number of the former graph is 356 and the latter is 333.

In 2004, the State Key Laboratory of Software Development Environment (SKLSDE) of Beihang University (also known as Beijing University of Aeronautics and Astronautics or BUAA) began a project named "Global IPv6 Networks Monitoring Project", one goal of which was similar to that of Scamper, i.e. to discover the accurate topology of the IPv6 Internet. But our approach was not simple traceroute. We noticed that the current tunnel linked way of global IPv6 ASes can be used to distribute the probing agents. Extending this idea, we designed the probing algorithm of the third party distribution and asynchronous traceroute. With these two new topology discovery approaches, we planted probing agents by configuring different tunnels instead of distributing the probe program to different geographic locations. Because of the great number of tunnels (up to Oct.2006, there was 91 usable tunnels in total), the probing data were enriched dramatically. The discovery system Dolphin [32] has been collecting data of routers, links, bandwidths and the AS statistics of the global IPv6 networks since 2004. In Mar. 2005 (the same time as that of the data which we got from Scamper), Dolphin discovered 3022 IPv6 addresses, 6825 links and 419 ASes. Up to Oct. 2006, 7401 IPv6 addresses, 51115 links and 508 ASes had been discovered from global backbone IPv6 networks. Comparing the discovered results by the two systems in TABLE 1, we can see that Dolphin's data are richer than Scamper. Besides, other statistics obtained by Dolphin are shown in TABLE 2 and more information is available on the website [32].

As a result, we have two data sources in this research: one is Scamper, and the other is Dolphin. The following section will make a comparison between the two AS topologies generated from the data sources which were obtained at the same time.

TABLE 1
Summary of the IPv6 Internet topology discovery by Scamper and Dolphin

|  | Scamper(Mar. 2005) | Dolphin (Mar. 2005） | Dolphin (Oct. 2006) |
| --- | --- | --- | --- |
| Number of probes | 17 | - | - |
| Number of ASes | 333 | 419 | 508 |
| Number of IPv6 Addresses | 2913 | 3022 | 7401 |
| Number of IP links | 7905 | 6825 | 51115 |

TABLE 2

More about Dolphin system

| Statistics | Values |
|---|---|
| Length of address list for probing | 5256 |
| Number of out-going packages per probing | 43,046,640 |
| Number of discovered ipv6 routers | 6325 |
| Number of discovered ipv6 address prefixes | 680 |
| Number of used tunnels | 91 |

3.2 Comparison between topologies from the two data sources

From the number of discovered ASes by Dolphin in TABLE 1, we can see the growth of the IPv6 Internet. Specifically, we discussed the data of March 2005 which is simultaneous with Scamper in the following.

In this section we compare the two topologies based on a list of graph metrics that have been found to be important in the networking literature. The list of metrics includes degree distribution, distance distribution, local clustering, normalized betweenness and rich-club connectivity as well as some basic graph properties such as numbers of nodes and edges, average degree and maximum degree. This list is not complete, but in this section's topology comparison and section 6's model comparison, we believed that those metrics are sufficiently diverse and comprehensive to be used as a good indicator of graph similarity [6]. The degree distribution $P(k)$ specifies the probability of nodes in a graph with the degree $k$. The distance distribution $D(d)$ is the number of pairs of nodes at the distance $d$, divided by the total number of pairs $n^2$ (self-pairs included). The local clustering $C(k)$ is the ratio of the average number of links between the neighbors of $k$-degree nodes to the maximum number of such links. The normalized betweenness is a weighted sum of the number of shortest paths passing through a given node or edge. The rich-club connectivity $\phi(r/n)$ is defined as the ratio of the number of links connecting the club members over the maximum number of allowable links $r(r-1)/2$ in which $r$ is the rank of a node denotes its position in a list of all nodes sorted in decreasing degree.

The topology from Scamper consists of 356 nodes and 1007 edges. And for the topology from Dolphin, 419 nodes and 1812 edges have been found. The basic graph properties and average values of the metrics are demonstrated in TABLE 3. More details are shown in Fig.1. The two topologies are very similar in the degree distribution (Fig.1a, b), the distance distribution (Fig.1c), rich-club connectivity (Fig.1d) and normalized node betweenness (Fig.1f) despite the differences of the average degree and the maximum degree. The local clustering distributions (Fig.1e) are close to each other for the small degrees.

The conclusion can be drawn that the discovered results by two systems agree well with each other in the distributions of degrees, distances, local clustering, normalized betweenness and rich-club connectivity. The results also indicate that the IPv6 Internet topology is a graph with high clustering, small mean-distance and obvious rich-club features.

TABLE 3
Comparison between the two AS topologies at the same time, March 2005, from the two systems

|  | **CAIDA Scamper** | **BUAA Dolphin** |
|---|---|---|
| Number of nodes | 356 | 419 |
| Number of edges | 1007 | 1812 |
| Avg degree | 5.657 | 8.649 |
| Exponent of $P(k)$ | 1.232 | 1.273 |
| Max degree | 85 | 119 |
| Avg distance | 2.934 | 2.783 |
| Nomalized avg node betweenness | 5.435E-3 | 4.255 E-3 |
| Nomalized avg edge betweenness | 2.914 E-3 | 1.535 E-3 |
| Avg clustering | 0.590 | 0.692 |
| Exponent of rich club | -1.169 | -1.268 |

## 3.3 The small degree exponent of the IPv6 Internet topology

The degree distribution is one of the most important metrics of networks topology. It is usually described on a frequency-degree log-log plot. If a network is scale-free, its degree distribution follows the power-law. On the plot, it is approximately linear and the slope corresponds to the degree exponent.

Despite the differences in the edge numbers and the maximum degrees, the topologies obtained by Scamper and Dolphin are quite similar in the degree distribution plots. The log-log plots of degree distribution of IPv6 topology from both Dolphin and Scamper are approximately linear, as shown in Fig.1a and Fig.1b. The correlation coefficient of linear regression is -0.9050 for Scamper and -0.9581 for Dolphin. Both are close to -1, which implies that their degree distributions follow the power-law. To be clearer, the complementary cumulative distribution plot is also given in Fig.1 g. As a result, it shows that the IPv6 Internet is scale-free at AS level.

But what is beyond expectation is that the exponent of degree distribution is only abut 1.2, far smaller than 2.2 of the IPv4 Internet [19, 20]. Since the data sources from Dolphin and Scamper are in good accordance with each other, it is not a coincidence that the result on the degree exponent of IPv6 is just 1.2. For the reason that the power-law of degree distribution is a main characteristic of a scale-free network, the big difference in the power-law exponent indicates that there is a new feature in the topology structure of the IPv6 Internet differing from the IPv4 Internet.

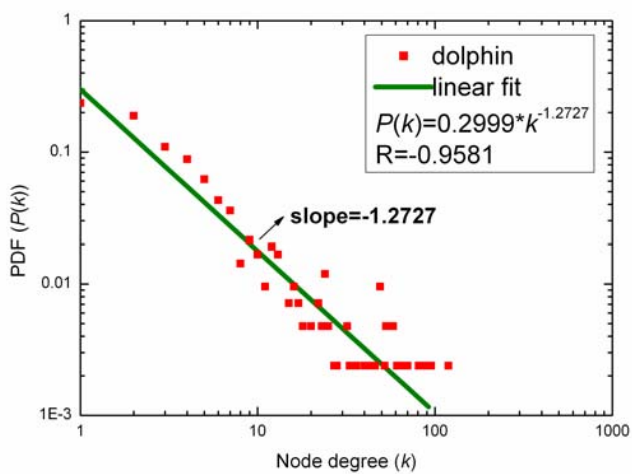

(a) Degree distribution from *dolphin*

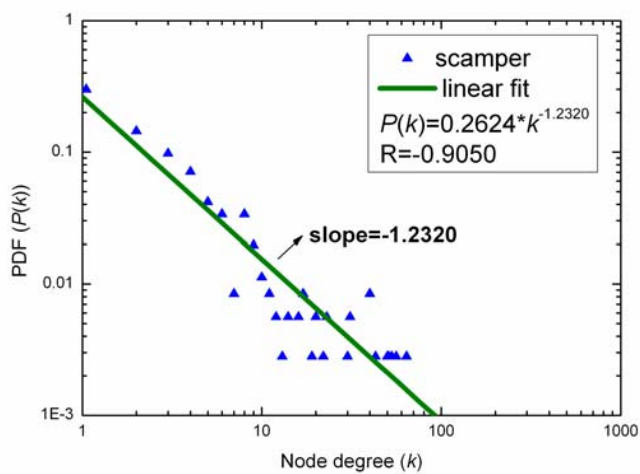

(b) Degree distribution from *scamper*

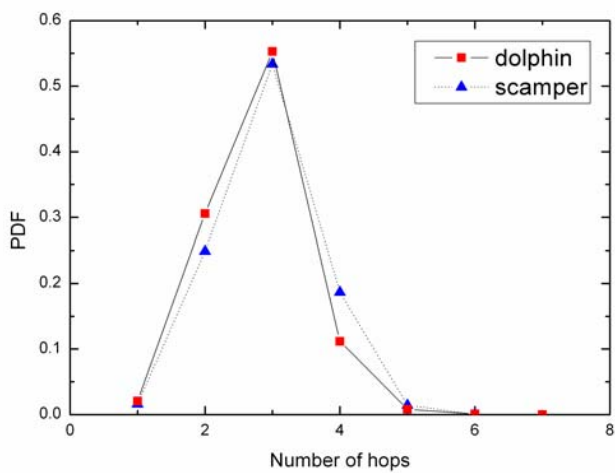

(c) Distance distribution

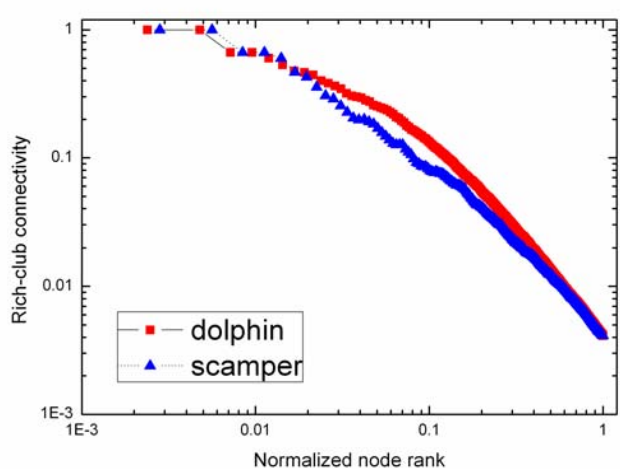

(d) Rich-club connectivity

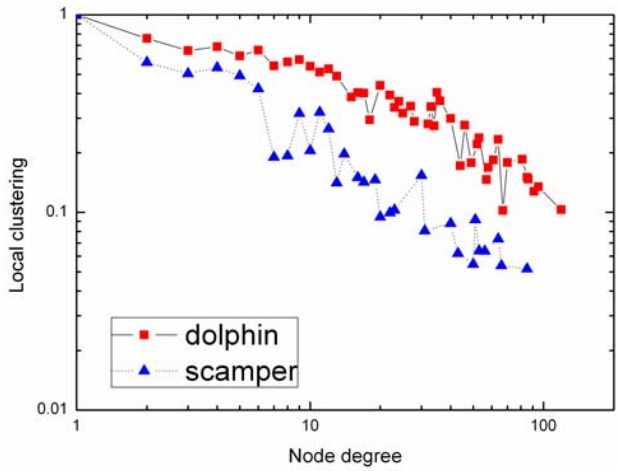
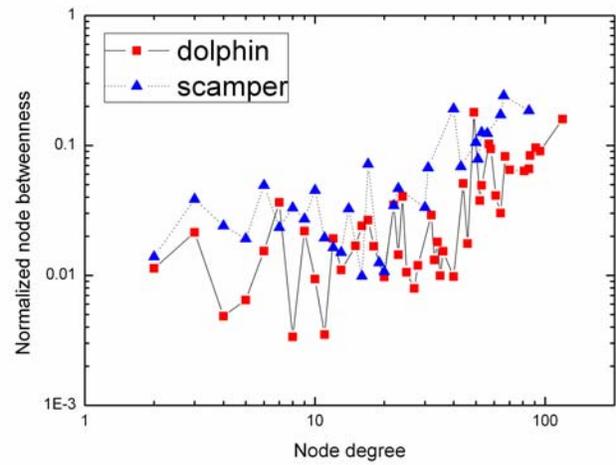

(e) Local clustering

(f) Normalized node betweenness

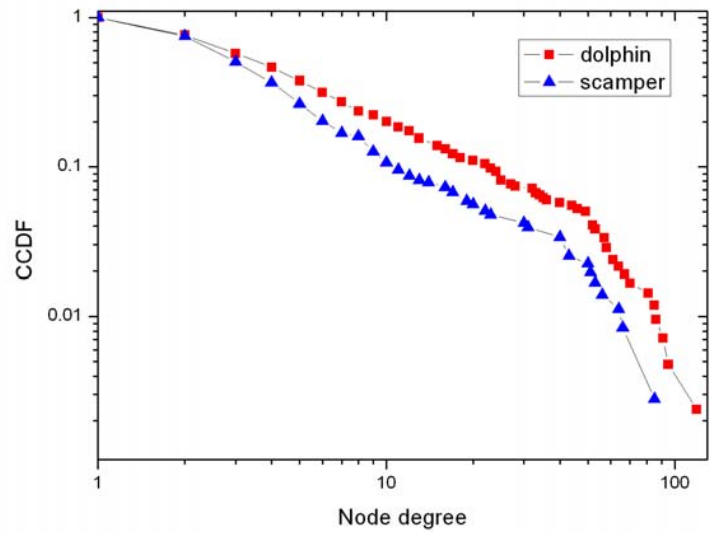

(g) Complementary Cumulative distribution of node degree

Figure 1. Comparison between the two topologies respectively discovered by *dolphin* system and *scamper* system

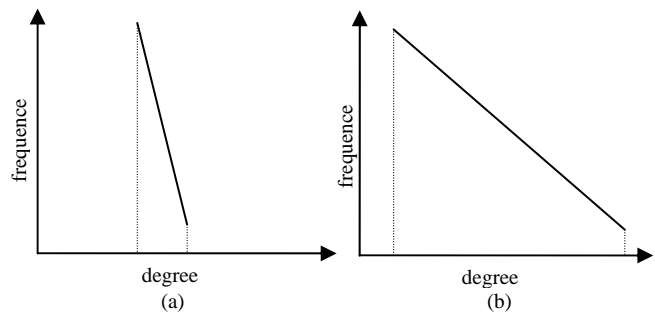

Figure 2. The degree exponent is a measure for topology uniformity of degree distribution. For networks of the same size, the exponent is bigger in (a), then its degree range is more concentrative and the topology is more uniform; while in (b), the exponent is smaller, therefore the degree range is wider and the topology is less uniform.

## 4. Explanation of small degree exponents and challenges for network evolving models

For degree distribution which follows the power-law, the slope of the regression line in a log-log plot can represent the degree exponent. For the networks of the same size, it is easy to obtain that if the degree exponent is bigger, then the degree range is more concentrative, as shown in Fig.2. Especially when the degree exponent is equal to infinity, all nodes have the same degree and it's a uniform topology. On the opposite side, for the smaller degree exponent, especially when the degree exponent is 0, all degrees have the same probability to appear in the topology, and thus the maximum degree can be much bigger than the minimum degree. Therefore, in general, the degree exponent is a measure of uniformity of the degree distribution. The bigger the exponent, the more uniform the topology; the smaller, the less uniform.

The IPv6 degree exponent is much smaller than that of IPv4, implying that the IPv6 Internet is less uniform than the IPv4 Internet. There are several possible reasons contributing to this. First of all, the size of current IPv6 Internet is much smaller than IPv4. Compared with more than 10 thousand ASes in the IPv4 Internet, less than 500 ASes of the IPv6 Internet constitute a rather small network. Then, IPv6 networks are on the stage of transition from the experimental use to practical global deployment, and for that reason, some characteristics of experimental networks are inevitable. For example, most of the large IPv6 internet experimental platforms are connected to almost every AS, while the smaller ASes are only linked to a few of those large experimental platforms (because no real application require the smaller ASes to directly connect to each other during the experimental period). Then the current IPv6 Internet resulted in the less uniform topology than that of the IPv4 Internet.

The degree exponents of most scale-free networks fall between 2 and 3 [7]. But besides the IPv6 Internet a small number of networks in other fields share the same feature of small degree exponents, such as the email network in social networks with the exponent of 1.5 and the soft package network in technology networks with the in-degree exponent of 1.4 and the out-degree of 1.6 [14]. The existence of scale-free networks with the degree exponent smaller than 2 is a reality. But most of the existing evolving models are only capable of constructing scale-free networks with the exponents bigger than 2, as shown in TABLE 4. When we wanted to reproduce the small degree exponent of the IPv6 Internet, we encountered a big challenge. In order to solve this problem, we extend the EBA model and propose a new model which has the capability to construct an evolving scale-free network with the degree exponent smaller than 2.

TABLE 4
The evolving models and their limits of the exponent $\gamma$

| Evolving network model | Features | Limits of $\gamma$ |
|---|---|---|
| BA model [8] | Linear growth, linear attachment | $\gamma = 3$ |
| Nonlinear attachment model [13] | $\Pi \propto k^\alpha$ only when $\alpha = 1$ can be scale-free | $\gamma > 2$ |
| Initial attractiveness model [33] | $\Pi \propto k + A$<br>$A$ is a constant attractiveness | $\gamma = 2$ if $A = 0$<br>$\gamma \to \infty$ if $A \to \infty$ |
| EBA model [9] | Edge-rewiring | $\gamma > 2$ |
| Accelerating growth model [17] | $\bar{k} = t^\theta$<br>*Directed* network | $\gamma = 3$ if $\theta \to 0$<br>$\gamma = 1.5$ if $\theta \to 1$<br>It's a directed network model. |
| Gradual aging model [17] | $\Pi(k_i) \propto k_i(t - t_i)^{-\nu}$ | $\gamma \to 2$ if $\nu \to -\infty$<br>$\gamma \to \infty$ if $\nu \to 1$ |
| PFP model [11] | $\Pi(k_i) \propto k_i^{1+\delta \lg k_i}$ | The experimental result shows $\gamma > 2$ |

This table is mainly based on Ref. [7].

## 5. Our model

As a measure for uniformity of degree distribution of scale free networks, the degree exponent is affected by many factors. Based on current evolving network models, we make use of two major factors to reproduce the feature of small degree exponent: the preferential attachment and the edge rewiring in network evolving process. For the preferential attachment, the linear relationship may be the most basic one. Although reality is far more complicated than that, statistics of many real networks show that the attachment probability is just a little higher than the linear preferential attachment [11]. Otherwise, if we change the preferential attachment probability as follows:

$$\Pi_i = \frac{k_i^{\alpha}}{\sum_j k_j^{\alpha}}, \tag{3}$$

it has been proven that the only case in which the topology of the network is scale free is that in which the preferential attachment is asymptotically linear. When $\alpha < 1$ it is a sub-linear attachment and the degree distribution is exponential; when $\alpha > 1$ it is super-linear attachment and almost all the links are attracted by one or two "super" node [13]. Our staring point is to carefully choose a linear attachment expression as follows:

$$\Pi_i = \frac{k_i + \varepsilon \cdot \overline{k}}{\sum_j k_j + \varepsilon \cdot \overline{k}}, \tag{4}$$

where $\Pi_i$ is the probability of node $i$ to obtain an edge, $k_i$ is the degree of node $i$, $\varepsilon$ is a constant, $\overline{k}$ is the average degree at the current time. If the value of $\Pi_i$ is negative after calculation, $\Pi_i$ will be set to zero to avoid being meaningless. The reason for picking $\overline{k}$ is that it can bring the features of the existing topology into the next stage of evolving process and it makes the theoretic calculation of the degree exponent much easier. The parameter $\varepsilon$ can adjust the node picked-up probability higher or lower than the pure linear preferential attachment in the BA model [8].

The edge rewiring can be understood as adjusting relationships between nodes. With it, the EBA model can generate a scale-free network with the degree exponent

$$\gamma = 1 + \frac{2m(1-q) + 1 - p - q}{m}. \tag{5}$$

In order to avoid getting exponential topologies, $p$ and $q$ must be restricted inside the scale-free regime. The EBA model can get good results of $\gamma > 2$ [9] and it makes clear that the edge rewiring mechanism is an important factor affecting the degree exponent.

Taking both the factors into consideration and based on the EBA model, we give out our model as follows:

We start with fully connected graph of $m_0$ nodes and at each time step we perform one of the following three operations:

(i) With probability $p$, we add $m(m<m_0)$ new links: we randomly select a node as the starting point of the new link, while the other end of the link is selected with probability given in (4).

(ii) With probability $q$, we rewire $m$ links: we randomly select a node and a link connected to it. Next we remove this link and replace it with a new link that connects with a new node chosen with probability given in (4).

(iii) With probability $1-p-q$, we add a new node: which has $m$ new links that with probability given in (4) connected to nodes already present in the system.

It is easy to see that the EBA model is a special case of our model with $\varepsilon = 1/\overline{k}$. For our model, the expression of the degree exponent is:

$$\gamma = 2(1-q)(1+\varepsilon) + 1. \tag{6}$$

In the following, we use the Continuum Theory [34] to prove the above result.

First, we assume that $k_i$ changes continuously. Consequently, the processes (i)–(iii) all contribute to $k_i$, each being incorporated in the continuum theory as follows.

(i) Addition of $m$ links with probability $p$:

$$\left(\frac{\partial k_i}{\partial t}\right)_{(i)} = \frac{pm}{N(t)} + pm\frac{k_i + \varepsilon\overline{k}}{M(t)}, \tag{7}$$

where $k_i(t)$ is the degree of the node $i$ at the time $t$, $N(t)$ is the number of all the nodes, at the time $t$ and $M(t) = \sum_{i=1}^{N(t)} (k_i(t) + \varepsilon \bar{k})$ ;

(ii) Rewiring of $m$ links with probability $q$:

$$\left(\frac{\partial k_i}{\partial t}\right)_{(ii)} = -\frac{qm}{N(t)} + qm\frac{k_i + \varepsilon \bar{k}}{M(t)} ; \tag{8}$$

(iii) Addition of a new node with probability $1-p-q$:

$$\left(\frac{\partial k_i}{\partial t}\right)_{(iii)} = (1-p-q)m\frac{k_i + \varepsilon \bar{k}}{M(t)} . \tag{9}$$

And at time $t$, we have:

$$N(t) = (1-p-q)t , \; M(t) = 2(1-q)(1+\varepsilon)mt , \tag{10}$$

$$\sum_{i=1}^{N(t)} k_i = 2(1-q)mt \text{ and } E = \varepsilon \bar{k} = \varepsilon \frac{\sum_{i=1}^{N(t)} k_i}{N(t)} = \frac{2\varepsilon(1-q)m}{1-p-q} . \tag{11}$$

By adding the contribution of the three processes, we obtain:

$$k_i(t) = (A + m + E)\left(\frac{t}{t_i}\right)^{\frac{1}{B}} - A - E , \tag{12}$$

where $A = \frac{2m(p-q)(1-q)(1+\varepsilon)}{1-p-q}$, $B = 2(1-q)(1+\varepsilon)$.

In the continuum theory, the probability $P(k_i)$ can be interpreted as the rate at which $k_i$ changes:

$$P[k_i(t) < k] = P\left[t_i > \left(\frac{m+A+E}{k+A+E}\right)^B\right] = 1 - P\left[t_i \leq \left(\frac{m+A+E}{k+A+E}\right)^B\right] = 1 - \left(\frac{m+A+E}{k+A+E}\right)^B \frac{t}{t+m_0} , \tag{13}$$

and we obtain:

$$P(k) = \frac{\partial P[k_i(t) < k]}{\partial k} = \frac{tB(m+A+E)^B}{t+m_0}(k+A+E)^{-B-1} . \tag{14}$$

Thus the degree distribution $P(k)$ follows a general power-law form and we get:

$$\gamma = B + 1 = 2(1-q)(1+\varepsilon) + 1 . \tag{15}$$

## 6. Model verifications

When analyzing statistics on the network power-law degrees, we adopt a usual method for topology analysis to capture "the power-law tail" better, which is to ignore some (no more than 5%) nodes whose frequencies are small but degrees are big[19, 20].

In our model the preferential attachment probability is $\pi(k) \propto k + \varepsilon\bar{k}$. To make sure it has physical meaning, $\pi(k)$ needs to be above zero. However in the case of $\varepsilon < 0$, if the percentage of nodes whose $\pi(k) < 0$ is less than 15%, the experimental results can still agree well with theoretical calculation after we make each node have at least one edge. And in the following, we carry out the experiments in this way.

The model verifications are carried out through a lot of experiments from three aspects: the verification of the degree exponent breaking the bound of 2, the verifications for different values of the exponent, and reproducing the AS topology of the IPv6 Internet.

6.1 The verification for breaking the bound of 2

Our model can generate a scale-free network with the degree exponent smaller than 2. The following parameters $q=0.525$, $\varepsilon=-0.25$, $p=0.3652$, $m=1$, $m_0=5$ were used to construct a large network of 100,000 nodes and 434,163 edges with the degree exponent $\gamma_e=1.7138$, shown in Fig.3. And the degree distribution follows the power-law with the correlation coefficient $R=-0.9441$ which indicates that the degree distribution of the topology is in good accordance with the power-law form. It is easy to see that the experimental value of the exponent $\gamma_e=1.7139$ also agrees well with the theoretical expected value $\gamma_t=1.7125$ according to (6).

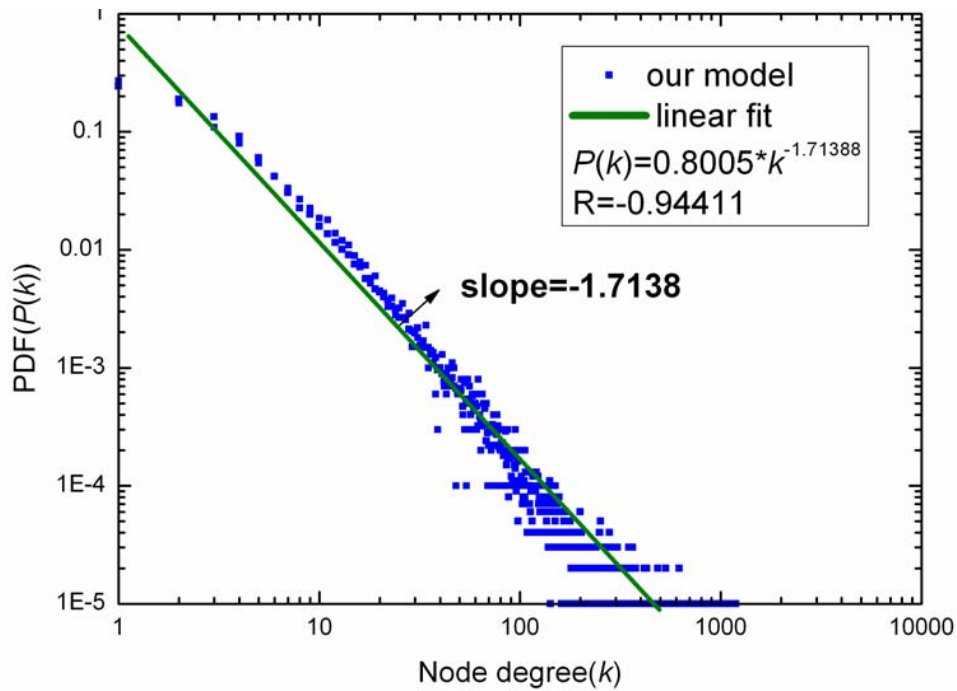

Figure 3. The degree exponent of the generated network, which is $\gamma_e=1.7138$, breaks the bound of 2 and is very close to the theoretical expected value $\gamma_t=1.7125$ according to (6).

6.2 The verifications for different values of the exponent

Our model is a generalization of the EBA model and the BA model. With $\varepsilon=1/\bar{k}$ our model is specialized to the EBA and with $\varepsilon=0$ and $q=0$ it is exactly the BA model. Not only can our model construct a network with the degree exponent smaller that 2 but also it has the capability to produce networks with different values of the exponent. Examples of such five networks with 10,000 nodes and 40,000 edges are shown in TABLE 5 and Fig.4 (a~e).

We noticed that in (6) $\gamma$ can be adjusted by $q$ and $\varepsilon$, so we did a series of experiments and then picked out a group with the biggest correlation coefficients. These experiments show that: firstly, the power-law correlation coefficient is almost close to 1, so the topology generated by this model is in good accordance with the power-law distribution; secondly, experimental results agree well with the theoretical expected value, as shown in TABLE 5.

TABLE 5
Five generated networks of the same size of 10,000 nodes and 40,000 edges

| No. | $q$ | $\varepsilon$ | $\gamma_t$ | $\gamma_e$ | $|R|$ | Figure |
|---|---|---|---|---|---|---|
| 1 | 0.525 | -0.25 | 1.7 | 1.712 | 0.951 | Fig.4a |
| 2 | 0.4 | -0.17 | 2.0 | 1.964 | 0.966 | Fig.4b |
| 3 | 0 | -0.25 | 2.5 | 2.480 | 0.976 | Fig.4c |
| 4 | 0 | 0 | 3.0 | 2.923 | 0.982 | Fig.4d |
| 5 | 0.2 | 0.56 | 3.5 | 3.443 | 0.987 | Fig.4e |

In this table, $q$ and $\varepsilon$ are the input parameters of each model; $\gamma_t$ is the theoretical expected value of the degree exponent by (6), $\gamma_e$ is the experimental result, and $|R|$ is the absolute value of the correlation coefficient.

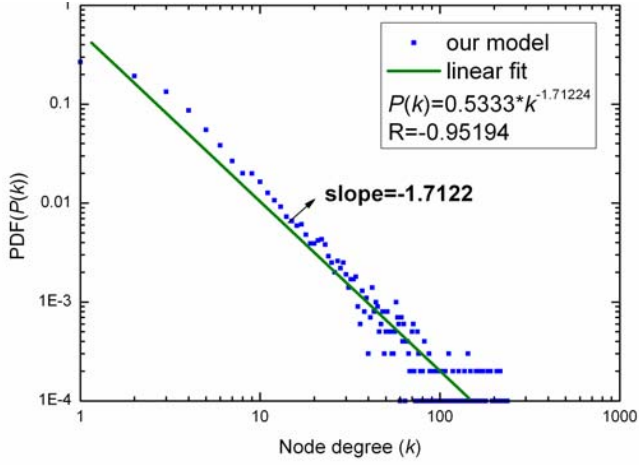

(a) $\gamma_t = 1.71$, $\gamma_e = 1.71$

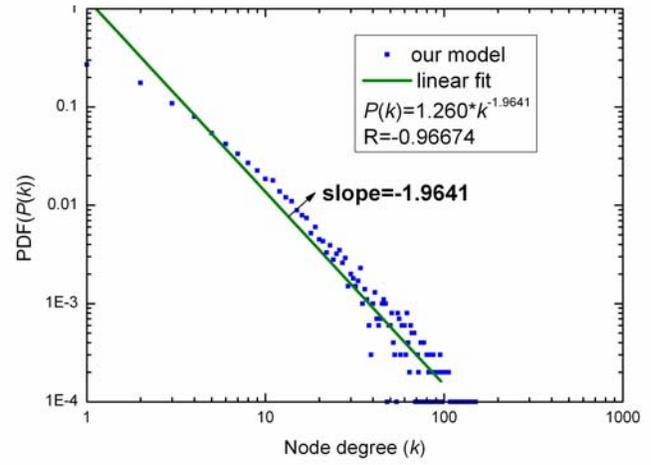

(b) $\gamma_t = 2.00$, $\gamma_e = 1.96$

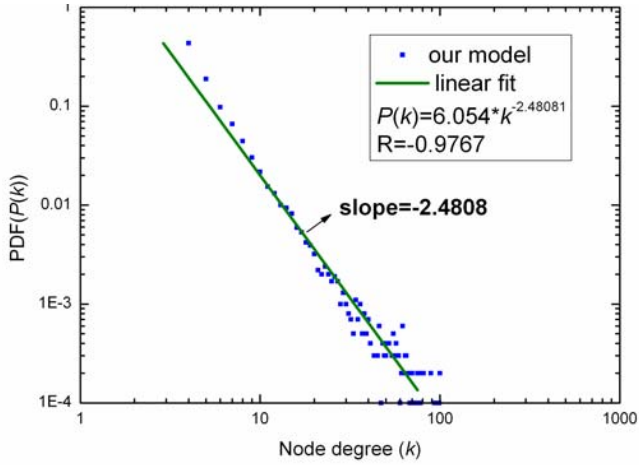

(c) $\gamma_t = 2.50$, $\gamma_e = 2.48$

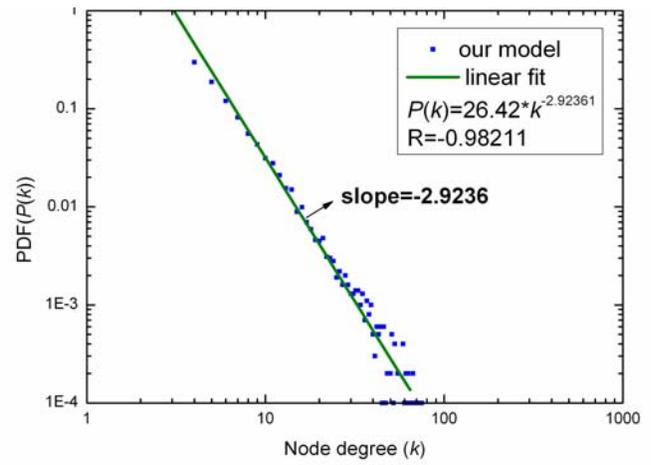

(d) $\gamma_t = 3.00$, $\gamma_e = 2.92$

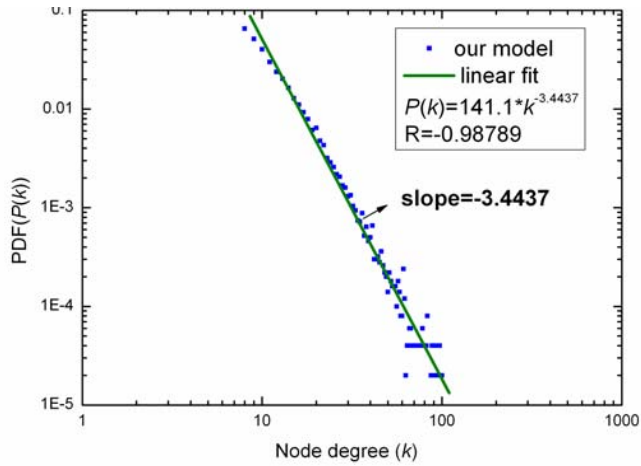

(e) $\gamma_t = 3.50$, $\gamma_e = 3.44$

Figure 4. Five networks of the same size of 10,000 nodes and 40,000 edges for different values of the degree exponent. In this figure, $\gamma_t$ is the theoretical expected value of the degree exponent by (6), and $\gamma_e$ is the experimental result.

6.3 Reproducing the IPv6 Internet AS topology

As mentioned in section 4, it is difficult to model the IPv6 Internet accurately because of its small degree exponent. In the following, we use our model to reproduce the IPv6 Internet AS topology discovered by Dolphin and the experimental results show that our model successfully capture the new feature of the small degree exponent as well as other topology metrics described in section 3.

The experiments show that our model has a convergence property. That is when the number of nodes grows to the infinity the degree exponent gradually reaches the theoretical value in (6). Therefore it can generate a scale-free network of small size whose degree exponent of degree distribution is smaller than the theoretical expected value in (6). Because the size of current IPv6 topology is relatively small, we can make use of this feature to generate a network. After a series of experiments, under this condition, i.e. $q=0.4$, $\varepsilon=-0.25$, $p=0.462$, $m=1$, and $m_0=5$, we generate a network with the topology almost the same as that of the IPv6 Internet as shown in TABLE 6 and Fig.5.

We also demonstrate the results by the BA and EBA models in TABLE 6 and Fig.5. In Fig.5a and Fig.5g, it's easy to see that the topology generated by our model closely match the IPv6 AS graph's degree distribution. And in Fig.5b, the linear regression shows our model successfully reproduce the small degree exponent of the IPv6 Internet AS topology. Fig.5(c~e) demonstrate that the reproduced topology is also in good accordance with the IPv6 Internet AS graph on metrics of the distance distribution, rich-club connectivity, node between-ness and local clustering.

TABLE 6
Reproducing the IPv6 Internet AS topology

|  | AS by Dolphin | BA Model | EBA Model | Our Model |
|---|---|---|---|---|
| Number of nodes | 419 | 419 | 419 | 419 |
| Number of edges | 1812 | 1661 | 1694 | 1771 |
| Avg degree | 8.64 | 7.92 | 8.09 | 8.45 |
| Exponent of $P(k)$ | 1.27 | 2.82 | 2.13 | 1.27 |
| Max degree | 119 | 55 | 57 | 96 |
| Avg distance | 2.78 | 3.00 | 3.12 | 2.75 |
| Nomalized avg node betweenness | 4.2E-3 | 4.7E-3 | 5.0E-3 | 4.2E-3 |
| Nomalized avg edge betweenness | 1.5E-3 | 1.8E-3 | 1.5E-3 | 1.6E-3 |
| Avg clustering | 0.692 | 0.041 | 0.293 | 0.476 |
| Exponent of rich club | -1.26 | -1.10 | -1.19 | -1.27 |

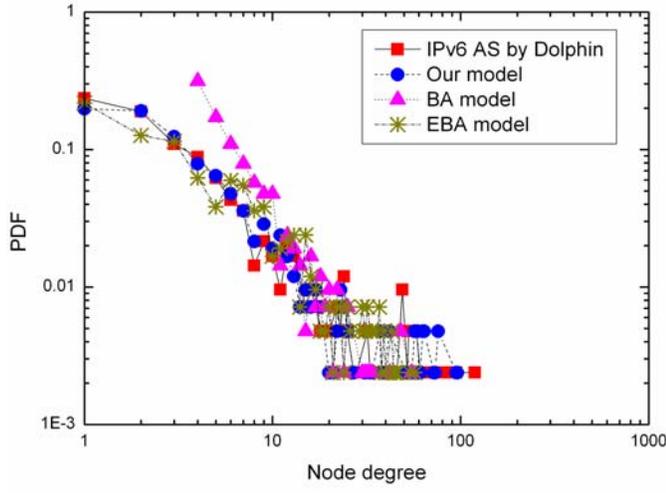

(a) Degree distribution comparison

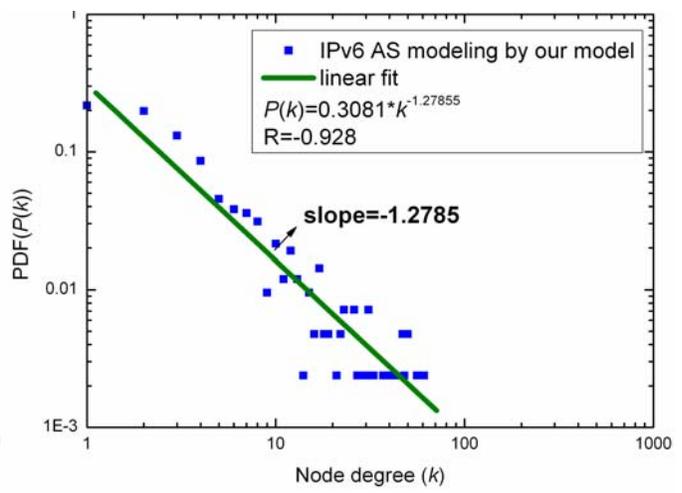

(b) Degree distribution linear regression

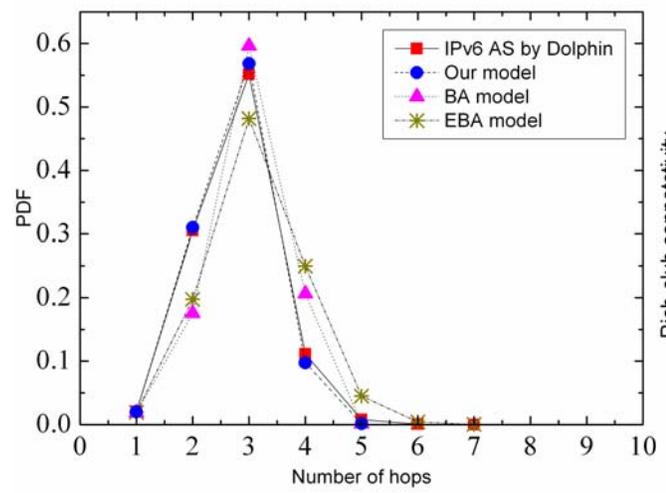

(c) Distance distribution

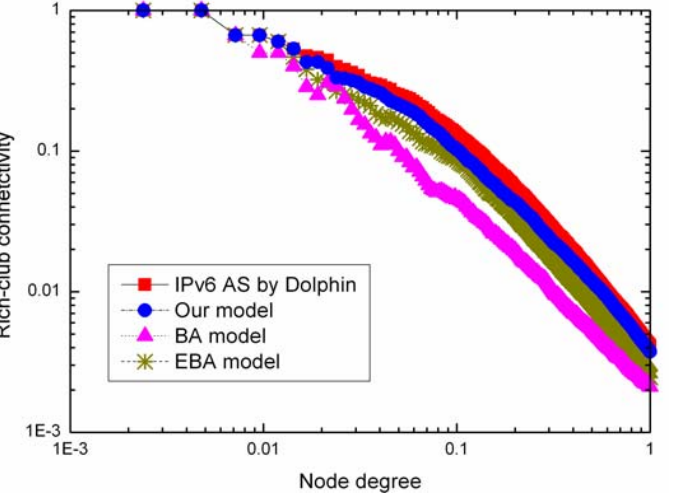

(d) Rich-club connectivity

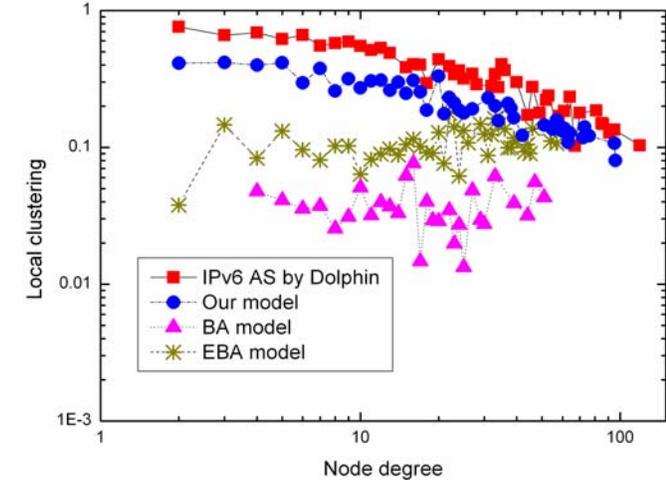

(e) Local clustering

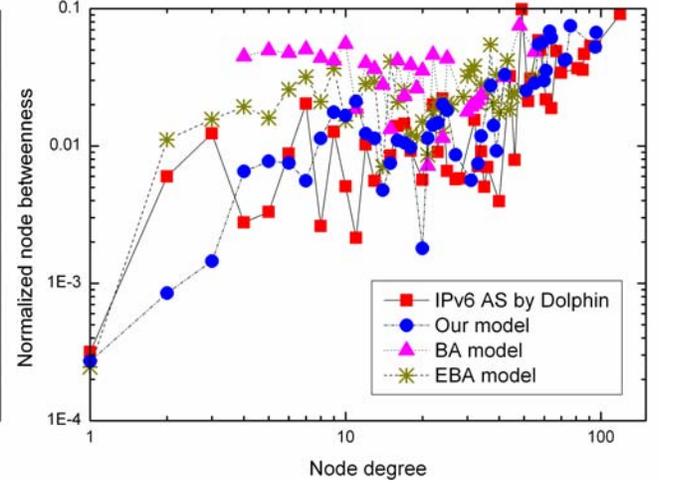

(f) Normalized node betweenness

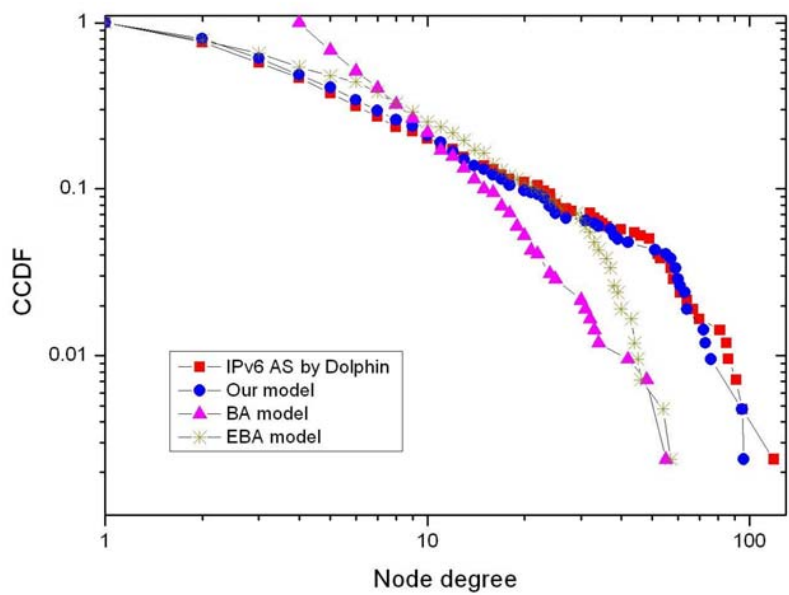

(g) Complementary Cumulative distribution

Figure 5. Reproducing the IPv6 Internet AS topology

## 7. Conclusion

Based on the analysis of the AS topologies of the IPv6 Internet from two data sources, we find that the IPv6 Internet is also a scale-free network but with a far smaller degree exponent than that of IPv4. As a measure of uniformity of the degree distribution, the degree exponent is affected by many factors. Based on the most important two of them: the probability of preferential attachment and edge rewiring in network evolving process, we propose our model which is a generalization of the EBA model. It is shown both theoretically and experimentally that this model breaks the bound of 2 for the degree exponent and it can well construct networks with different values of the degree exponent. Finally using this model we successfully reproduce the topology of the IPv6 Internet.

**Acknowledgements**

The authors would like to thank Professor Guanrong Chen for his helpful comments and suggestions and Matthew Luckie for his help on Scamper data analysis and CAIDA for the topology data of Skitter and Scamper. We are also very grateful to the anonymous reviewers whose valuable comments and constructive criticism helped to improve the paper significantly.